# Reproducing some observed galactic rotation curves without dark matter or modified Newtonian dynamics

Jau Tang[1*] and Qiang Tang[2*]

[1]Wuhan National Laboratory for Optoelectronics,

Huazhong University of Science and Technology, Wuhan 430074, China

[2]School of Artificial Intelligence,

Anhu University of Science and Technology, Huainan, Anhui 232000, China

J. Tang:   jautang@hust.edu.ch; Q. Tang:   2022135@aust.edu.cn

**Abstract**

Dark matter has been a long-standing and important issue in physics, but direct evidence of its existence is lacking. This work aims to elucidate the mystery and show that the dark matter hypothesis is unnecessary. We can nicely reproduce the observed rotation curves using only conventional Newtonian dynamics based on experimental surface brightness profiles of several galaxies. Our success is based on realizing that the mass radial distribution follows a stretched exponential decay with a small exponent over a few hundred kiloparsecs. Our quantitative analysis indicates that for these four example galaxies, there is no need to invoke the hypothetical dark matter presently unknown to humans or the modified Newtonian dynamics (MOND) paradigm.

## 1. Introduction

According to Newton's theory of gravitation, the gravitational force is inversely proportional to the square of the distance between two objects. In our solar system, with the sun containing 99.8% of the mass of the whole solar system, the rotation velocity for a planet at a distance $R$ from the sun follows Kepler's law of $v = \sqrt{GMsun/R}$. However, the rotation curve of orbiting stars in a spiral galaxy has been known to deviate significantly from Kepler's relation. Such deviation is not surprising because the density distribution is not concentrated at the galactic center. Since the earlier works of Oort, Zwicky in the 1930s[1] and Rubin[2] in the 1960s, astronomers have become aware of the missing matter in galaxies[3]. The rotation of galaxies appears to be too fast, and the gravitational pull from the masses of all visible constituents seemingly is insufficient to maintain a stable rotating galaxy without its spiral arms being ripped apart. Therefore, the dark matter hypothesis has become the mainstream paradigm to account for the missing matter[4]. It has been widely estimated that the universe consists of about 4% ordinary matter, 26% dark matter, and the remaining 70% dark energy [5]. It is commonly believed that a cold dark matter halo has a spherical radius of ~100 kpc from the galactic center. Direct evidence for such exotic dark matter has remained elusive, despite extensive experimental efforts by astronomers and high-energy physicists. Consequently, it is generally believed that dark matter must be drastically different from all known matter to humankind, and its physical properties are believed to be beyond the scope of the Standard Model[6,7]. As an alternative to solve the missing matter problem, MOND (Modified Newtonian Dynamics)[8-10] was proposed in the 1980s[10]. Although such a hypothesis could explain the observed rotation curves, it overthrows the well-established Newton's inverse-square force law and Einstein's general relativity theory of gravitation. To most scientists, MOND is an ad hoc fix to the problem but lacks fundamental physical understanding and has several serious issues explaining other facts. In contrast, our analysis in this work relies only on the gravitation theory of Newton and the experimentally observed galactic surface brightness profiles. With these, we are able to quantitatively reproduce the hallmark rotation-curve characteristics, showing an initial rise in velocity and then a leveling off at increased distances. Our analysis implies that there is no need to invoke the dark matter or MOND hypothesis to explain the rotation curves of galaxies.

## 2. Theory and results

To analyze the rotation curve of a galaxy, we first consider, for simplicity, a density distribution $\rho(r)$ with spherical symmetry. A more realistic disk-like density profile using an ellipsoid-shaped model will be examined later. The Newtonian gravitational potential energy $U(R)$ of an orbiting star at a distance R is given by

$$U(R) = 2\pi G \int_0^\infty dr \rho(r) r^2 \int_0^\pi d\theta \sin\theta \frac{1}{\sqrt{r^2 \sin^2\theta + (R - r\cos\theta)^2}}. \tag{1}$$

For a spherical density distribution, the above equation can be reduced to

$$U(R) = \frac{4\pi G}{R} \int_0^R dr\, r^2 \rho(r) r^2 + 4G \int_R^\infty dr\, r \rho(r). \tag{2}$$

For an orbiting star with a velocity $v(R)$ in a stable orbit, its centrifugal force must be balanced by the gravitational force from the galaxy, and one has

$$v^2(R) = -R \frac{d}{dR} U(R). \tag{3}$$

Consequently, one has the rotation curve as

$$v(R) = \left(\frac{G}{R} \int_0^R 4\pi dr\, r^2 \rho(r)\right)^{1/2}. \tag{4}$$

The above equation for the rotation curve is rigorously derived from Eqs. (1-3), which are entirely based on the Newtonian dynamics. In our solar system, where the density distribution is concentrated at the sun with a mass M, $\rho(r)$ behaves like a Dirac $\delta$- function. Consequently, Eq. (4) leads to Kepler's law of $v(R) = \sqrt{GM/R}$. Generally speaking, Eq. (4) indicates that the density integration solely determines the velocity at R from 0 to R, namely, the orbiting velocity is independent of the mass distribution outside the sphere of R. In most galaxies, the reported rotation curve covers a span of about 30 kpcs, and typically shows an initial rise within 5 to 10 kpc before reaching a plateau. Therefore, the density distribution at distances outside the integration sphere does not contribute to the rotation velocity, regardless of whether the mass is from ordinary matter or dark matter.

Unlike our solar system, the density distribution in most galaxies is widely spread, and the rotation curve is expected to deviate from Kepler's law significantly. We now consider a density distribution with a stretched exponential dependence as $exp(-(R/R_0)^s)$, where s is an exponent and $R_0$ is a characteristic length. Using Eq. (4) we calculate the corresponding rotation curve for several values of the exponent s from 2 to 0.25. We found that the rotation curve for a larger s exhibit an initial rise for the orbiting velocity, and as $R/R_0$ increases a drop-off appears sooner for a larger s. As s decreases the curve becomes to show a flat top after peaking. Such a functional form has been used in analyzing the galactic surface brightness profile $\Sigma_L(r)$. Such a profile, known as the de Vaucouleurs profile[11] with s = ¼, and the Sérsic profile[12] with an arbitrary s, is related to a galaxy's mass density profile. In our model we consider a linear dependence of the integrated luminosity to the total mass of the galaxy. If the dependence is not exact linear, so long as the surface brightness curve follows a stretched exponential with a small exponent = close to 0.5 or smaller, we can reproduce the same rotation curve with a flat top., although he exact velocity on the plateau could vary.

To calculate the rotation curve from Eq. (4), one must first determine the density distribution based on the surface brightness profile which the Sérsic model12 c with $\Sigma_L(r) \equiv \Sigma_0 \exp(-b(s)(r/R_0)^s)$. The deprojection of the Sérsic profile to a 3D luminosity density function $\rho_L(R)$ can be obtained by[13,14]

$$\rho_L(R) = -\frac{1}{\pi} \int_R^\infty dr \frac{1}{\sqrt{r^2 - R^2}} \frac{d}{dr} \Sigma_L(r) \tag{5}$$

It can be shown[12] that $\rho_L(R)$ follows approximately the same stretched exponential function, except with an additional power-law function multiplier, as given in the following

$$\begin{aligned}
&\rho_L(R) \approx \rho_{L,0} (R/R_0)^{-p(s)} \exp\left(-b(s)(R/R_0)^s\right) \\
&p(s) \approx 1 - 0.6097\,s + 0.05463\,s^2 \\
&b(s) \approx 2/s - 1/3 + 0.00986\,s + 0.0018\,s^2 + 0.000114\,s^3 - 0.0000715\,s;
\end{aligned} \tag{6}$$

and the above approximation holds for $1.79 \geq s \geq 0.1$ and $3 > \log(R/R_0) > -2$. As shown above, the 3D mass density distribution also follows a stretched exponential dependence with a power-law multiplier. For a comparison with the results in FIG. 1, we now consider a surface brightness profile $\Sigma(r)$ defined as $\exp(-(R/R_0)^s)$ with s = 2, 1, 0.6, 0.5, 0.4, and 0.3 as well. Here we calculate the 3D luminosity density profile $\rho_L(R)$ by numerical integration in Eq. (5), which is related to the mass density profile by a proportional constant. The calculated rotation curves, as illustrated in Fig. 1, all appear to have an initial rise at a smaller $R/R_0$. The curve peaking for a large s appears to occur at a distance closer to the galactic center, and the level-off for a small s also occurs at a smaller $R/R_0$. This analysis again provides a clear clue that the key to explaining the observed rotation curves is the stretched exponential profile with a small s, and not cold matter halo.

To obtain a quantitative analysis of the actual surface brightness profiles and the rotation curves, we extracted the data points from the experimental surface brightness profiles of several galaxies reported in the literature.[15] The data curves of four galaxies (Abell 2029, Abell 2199, Abell 1413, and Abell 1759) were chosen because these curves are rather smooth and cover a wide range from a few pics to a few hundred of kpcs. To improve the overall fit to the data points in the whole range from short to long distances, we used a least-square fit procedure to fit the extracted data using the same logarithmic scale by a sum of two stretched exponential functions

$$\Sigma_L(r) = A_1 \exp(-F_1 r^{s_1}) + A_2 \exp(-F_2 r^{s_2}). \qquad (7)$$

The mass density profile $\rho_L(R)$ is related to the luminosity profile by a proportional constant, i.e., $\rho(R) = \alpha \rho_L(R)$. By definition, the total luminosity and the proportional constant are related to the galaxy's overall mass $\alpha = M_{sum}/L_{sum}$. According to the literature[16-18], the total mass for Abell 2029, Abell 2199, Abell 1413, and Abell 1795 are given by 4.1 x10$^{14}$, 3.2 x10$^{14}$, 3.1 x10$^{14,}$ and 2.7x10$^{14}$ M☉, respectively. The fitted curves are shown in Fig. 2, and the relevant fit parameters are given in Table 1.

With these parameters, we can use Eq. (5) to calculate the 3D luminosity density profile and an approximate Eq. (4) to calculate the rotation curve $v(R)$. The mass density profile is related to the luminosity profile by a proportional constant. Consequently, the rotation curve can then be calculated by the following equation

$$v(R) = \left(-\frac{G}{\pi R}\left(\frac{M_{sum}}{L_{sum}}\right)\int_0^R dr 4\pi r^2 \int_r^\infty dr_1 \frac{1}{\sqrt{r_1^2 - r^2}} \frac{d}{dr_1} \Sigma_L(r_1)\right)^{1/2}$$

$$= \sqrt{\frac{\alpha G}{\pi}} \left(-\frac{1}{R}\int_0^R dr r^2 \int_r^\infty dr_1 \frac{1}{\sqrt{r_1^2-r^2}} \frac{d}{dr_1} \Sigma_L(r_1)\right)^{1/2} \qquad (8)$$

where $G = 4.3 \times 10^{-6} kpc$ M☉ (Km/s)$^2$, $r$ is in kpc, and the values for all relevant fit parameters are given in Table 1. The proportional constant for each galaxy is averaged to obtain α$_{ave}$ in km/s. The A-type dash=;-ne curves in Fig. 3 illustrate the calculated rotation curves based on the surface brightness data. The curves for these four galaxies exhibit an initial rise and slowly approach a flat top at a distance greater than 50 kpcs. The curve pattern can be improved to have a steeper rise and quickly reaches a flat top at a much shorter distance within 10 kpsc if an extra mass density distribution is included near the center. By including such an extra mass distribution for the bulge which can be represented by

$$\rho_B(R) = A_0 \, R^{-p(s)} \exp\left(-b(s)(R/R_0)^s\right), \tag{9}$$

as in Eq. (6), except our re-defining the constant coefficient $A_0$. This inclusion significantly improves the rotation curve pattern at short distances. Using $A_0$, $R_0$ and $s_0$ given in Table 2 for this additional density profile $\rho_B(R)$ from the bulge, the resultant rotation curves are illustrated by the B-type solid-line curves in Fig. 3. These curves show a steep rise to approach a flat plateau. Such patterns closely resemble the experimental rotation curves of most galaxies. In our model calculations, we only need the experimentally observed surface brightness profile from luminous objects, plus an additional mass distribution inside the galactic bulge. This extra mass added in our model is rather small, about 1~3 % of the total mass of the whole galaxy. Such an additional source near the galactic bulge (with a typical radius of ~ 3 kpcs), could come from non-luminous or less luminous massive objects, such as supermassive black holes, brown dwarfs, neutron stars, etc. that are hindered from contributing to the observed surface brightness profile. In addition, the surface brightness data curves generally do not contain sufficient data points at short distances, thus the luminosity profile does not sufficiently represent the actual mass distribution inside the bulge.

So far, the above analysis and simulations are based on the simple spherical model for the galactic density distribution. Because the galactic shape is generally non-spherical, we need to consider such anisotropy in the analysis. For simplicity, we assume an ellipsoidal density profile with a short axis along y and equal-length long axes along x and z. To treat galactic rotation curves, we consider the following stretched exponential radial dependence with an exponent s for the density profile as

$$\rho(x,y,z) = \rho(0) \, exp\left(-\left((x^2 + z^2 + y^2/\eta^2)/R_0\right)^{s/2}\right).\text{s} \tag{10}$$

Here we consider the short axis of the ellipsoid is along the y axis, and the flatness parameter $\eta$ is the ratio of the short-axis length and the long-axis length. The parameter $\eta$ controls the flatness of the ellipsoid, e.g., for a sphere $\eta = 1$ and as $\eta$ decreases the ellipsoid appears like a disk. For a typical galaxy, $\eta$ is on the order of 0.1. Defining $x = x_1, y = y_1 \eta, z = z_1$ and at an observation point along the z-axis at a distance $R$ to the galactic center one has the gravitational potential as

$$U(R) = G\rho(0)\eta \int_{-\infty}^{\infty} dx_1 \int_{-\infty}^{\infty} dy_1 \int_{-\infty}^{\infty} dz_1 \frac{e^{-F(x_1^2+y_1^2+z_1^2)^{s/2}}}{\sqrt{x_1^2+\eta^2 y_1^2+(R-z_1)^2}}$$

$$= 4Gb\rho(0)\eta \int_0^{\infty} dr\, r^2 e^{-(r/R_0)^s} \int_{-1}^{1} d\xi \int_0^{\pi/2} d\phi \frac{1}{(r^2+R^2-2Rr\xi-r^2(1-\eta^2)(1-\xi^2)\sin^2\phi)^{1/2}}. \quad (11)$$

The above expression for an ellipsoidal density profile is more general than that in s Eqs. (1) and (2) for a spherical profile. According to the Newtonian dynamics, based on the above gravitation potential but in the spherical coordinate system we obtain the following relation for the square of the rotation velocity

$$v^2(R) = G\rho(0)\eta \int_0^{\infty} dr\, r^2 e^{-Fr^s} \int_{-1}^{1} d\xi \int_0^{2\pi} d\varphi \frac{R(R-r\xi)}{\left(r^2+R^2-2Rr\xi-r^2\left(1-\eta^2\right)\left(1-\xi^2\right)\sin^2\varphi\right)^{3/2}} \quad (12)$$

For the simple case of $\eta = 1$ for a spherical density profile, the above equation can be simplified to Eq. (4) by integration of $\phi$ and $\xi$. If the density profile $\rho(R)$ for a disk-like ellipsoid-shaped galaxy is composed by several stretched exponential radial functions, Eq. (8) for $v(R)$ of the spherical case needs to be replaced for the ellipsoidal case by

$$v^2(R) = G\left(\frac{M_{sum}}{L_{sum}}\right) \int_0^{\infty} dr\, r^2 \frac{\rho(r)}{\rho(0)} \int_{-1}^{1} d\xi \int_0^{2\pi} d\varphi \frac{R(R-r\xi)}{\left(r^2+R^2-2Rr\xi-r^2\left(1-\eta^2\right)\left(1-\xi^2\right)\sin^2\varphi\right)^{3/2}} \quad (13)$$

or

$$v(R) = \sqrt{\alpha G \int_0^\infty dr\, r^2 \frac{\rho(r)}{\rho(0)} \int_{-1}^1 d\xi \int_0^{2\pi} d\varphi \frac{R(R-r\xi)}{\left(r^2+R^2-2Rr\xi-r^2(1-\eta^2)(1-\xi^2)\sin^2\varphi\right)^{3/2}}} \qquad (14)$$

where $\alpha = M_{sum}/L_{sum}$.

Using Eq. (14) we calculated the rotation curves for two galaxies, Abell 2029 and Abell 2199, as shown in Fig. 4 to illustrate the effect of the parameter $\eta$. This flatness parameter controls the shape of the ellipsoid, for $\eta = 1$ the shape is spherical as we have analyzed earlier, but as $\eta$ becomes smaller the shape approaches a flat disk. For a typical spiral galaxy, one has $\eta \sim 0.1$. It can be seen from Fig. 4 that the plateau velocity for $\eta = 0.1$ is greater than that of $\eta = 1$. It indicates that, due to more dense mass distribution on the disk plane, the effective gravitation pull for a disk is stronger -like galaxy than a spherical galaxy. Here, we compare the experimental velocity at the flat top and our calculated value. As shown in Fig. 4a, for Abell 2029 the observed plateau velocity[19] is about 470(+/-60) km/s which is very close to our calculated value of 475 km/s for a flat ellipsoidal galaxy with $\eta = 0.1$. In comparison, the plateau velocity is about 385 km/s for a spheroidal case with $\eta = 1$. Based on Fig. 4, the simulations indicate that one needs ~ 2 % of the overall galaxy mass for the extra mass in the bulge to produce the desired steep rise for the orbiting velocity at very short distance.

## 3. Discussion and conclusions

We present in this report an analysis of the galactic rotation curve, solely based on the Newtonian dynamics and the experimental surface brightness profile. We nicely reproduce the observed galactic rotation curves without the dark matter hypothesis or MOND paradigm to explain this phenomenon. Using quantitative analysis of the rotation curves for several galaxies, we disprove the necessity of these hypothetic paradigms. We show in Fig. 2 that the surface brightness profiles of these galaxies all follow a stretched exponential shape of $exp(-(R/R_0)^s)$. By least-square fit to the experimental data of four galaxies (Abell 2029, Abell 2199, Abell 1413, and Abell 1795), we obtained a fit value for the exponent s ~ 0.5. Moreover, our calculated curves in Fig. 4 quantitatively reproduce the observed pattern with an initial rise followed by a flat top. Thus, as far as the galactic rotation-curve issue is concerned, our analysis raises a critical question about whether dark matter is real or not.[22] In our model we assume a linear dependence of the integrated luminosity on the total mass of the galaxy. If not so, so long as the surface brightness curve follows a stretched exponential with a small exponent about 0.5 or smaller, we can reproduce the desired rotation curve as observed, showing a sharp rise and then a flat top for the rotation velocity as the distance increases. Although the velocity on the plateau could vary, the desired shape remains unchanged without the hypothetical assumption of dark matter or MOND.

Now after we have settled the issue about whether dark matter is needed or not for the explanation of the galactic rotation curve, one may wonder how one could describe the gravitational lensing which is commonly attributed to light bending in the curved space-time induced by dark matter.[20] Here, we offer another gravitational lensing mechanism that involves gravitation-induced variations of the refractive index for the interstellar media. Refractive media gradients, such as very thin gases or dust, could exist throughout the interstellar space within several hundred kpcs or more around a galactic center, due to a slow decaying stretched exponential density profile. Bending of light could occur due to a gravitation-induced refractive-index gradient[21], but this refractive medium mechanism does not cause bending of neutrinos which experience no electromagnetic interactions. Therefore, one could compare gravitational lensing between photons and neutrinos to determine which mechanism is operating. Further analysis is needed to answer whether light bending is solely due to Einstein's curved space-time effects or could be caused by the gravitation-induced density variations of the interstellar refractive media. In short, as far as the galactic rotation curve is concerned, our reproduction of the rotation curve does not rely on the dark matter hypothesis which assumes the existence of exotic particles outside the scope of the Standard Model. Our analysis is consistent with the Newtonian dynamics and rules out the necessity of the hypothetic MOND paradigm that deviates from the well-trusted gravitation theory of Newton or Einstein.

TABLE 1 | Fitted parameters to the surface brightness profiles of four galaxies using two stretched exponential density functions and the total mass from both components $M_{1+2}$ in the solar mass $M_\odot$.

|  | A2029 | A2199 | A1413 | A1795 |
|---:|---|---|---|---|
| $A_1$ | 1.37x10$^{-3}$ | 5.85x10$^{-3}$ | 3.37x10$^{-4}$ | 1.44x10$^{-3}$ |
| $A_2$ | 1.22x10$^{-3}$ | 5.69x10$^{-5}$ | 1.03x10$^{-4}$ | 1.00x10$^{-4}$ |
| $F_1$ | 1.20x10$^{-1}$ | 5.51x10$^{-1}$ | 1.65x10$^{-1}$ | 2.15x10$^{-1}$ |
| $F_2$ | 6.83x10$^{-2}$ | 5.90x10$^{-2}$ | 5.50x10$^{-2}$ | 4.10x10$^{-1}$ |
| $S_1$ | 0.65 | 0.45 | 0.542 | 0.56 |
| $S_2$ | 1.20 | 1.42 | 1.25 | 1.31 |
| Liam | 19.4 | 12.7 | 13.4 | 13.6 |
| $M_{sum}$ | 8.0 x10$^{14}$ | 6.5 x10$^{14}$ | 7.6 x10$^{14}$ | 6.0x10$^{14}$ |
| $\alpha$ | 4.1x10$^{13}$ | 5.1 x10$^{13}$ | 5.7 x10$^{13}$ | 4.4 x10$^{13}$ |
| $\sqrt{\alpha_{ave}G/\pi}$ | 7.5 x10$^{3}$ | 7.5 x10$^{3}$ | 7.5 x10$^{3}$ | 7.5 x10$^{3}$ |

TABLE 2 | Fitted parameters obtained from Table 1 plus the parameters for $A_0$, $R_0$ and $s_0 = 1.5$ to model the additional short-distance contribution from the galactic bulge.

|  | A2029 | A2199 | A1413 | A1795 |
|---|---|---|---|---|
| $A_1$ | $1.37 \times 10^{-3}$ | $5.85 \times 10^{-3}$ | $3.37 \times 10^{-4}$ | $1.44 \times 10^{-3}$ |
| $A_2$ | $1.22 \times 10^{-3}$ | $5.69 \times 10^{-5}$ | $1.03 \times 10^{-4}$ | $1.00 \times 10^{-4}$ |
| $F_1$ | $1.20 \times 10^{-1}$ | $5.51 \times 10^{-1}$ | $1.65 \times 10^{-1}$ | $2.15 \times 10^{-1}$ |
| $F_2$ | $6.83 \times 10^{-2}$ | $5.90 \times 10^{-2}$ | $5.50 \times 10^{-2}$ | $4.10 \times 10^{-1}$ |
| $S_1$ | 0.65 | 0.45 | 0.542 | 0.56 |
| $S_2$ | 1.20 | 1.42 | 1.25 | 1.31 |
| $A_0$ | $1.10 \times 10^{-3}$ | $9.94 \times 10^{-4}$ | $1.7 \times 10^{-4}$ | $7.2 \times 10^{-4}$ |
| $R_e$ | 3.5 kpc | 4.0 kpc | 5.0 kpc | 4.5 kpc |
| $M_0/M_{sum}$ | 1.5% | 2.8% | 1.0% | 2.6% |

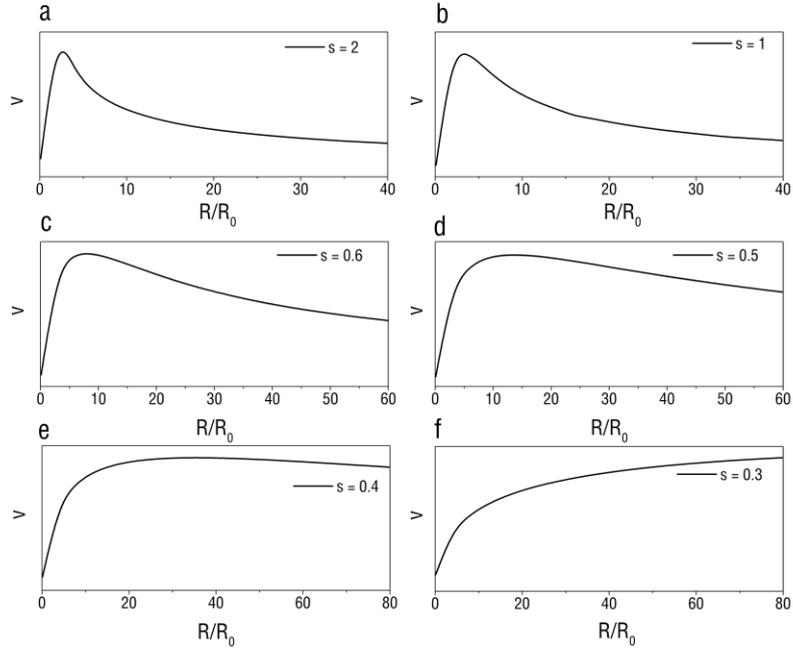

Fig. 1 Calculated rotation curves from EQ. (4) and (5) as a function of $R/R_0$. We assume a surface brightness profile of $exp(-(R/R_0)^s)$ for s = 2, 1, 0.6, 0.5, 0.4 and 0.3. For a large s = 2 or 1, such as for a Gaussian or simple exponential decay, the curve shows peaking then dropping off, whereas the curve appears to have a flat top as s becomes smaller. This analysis indicates that the rotation-curve shape is dictated solely by the s parameter.

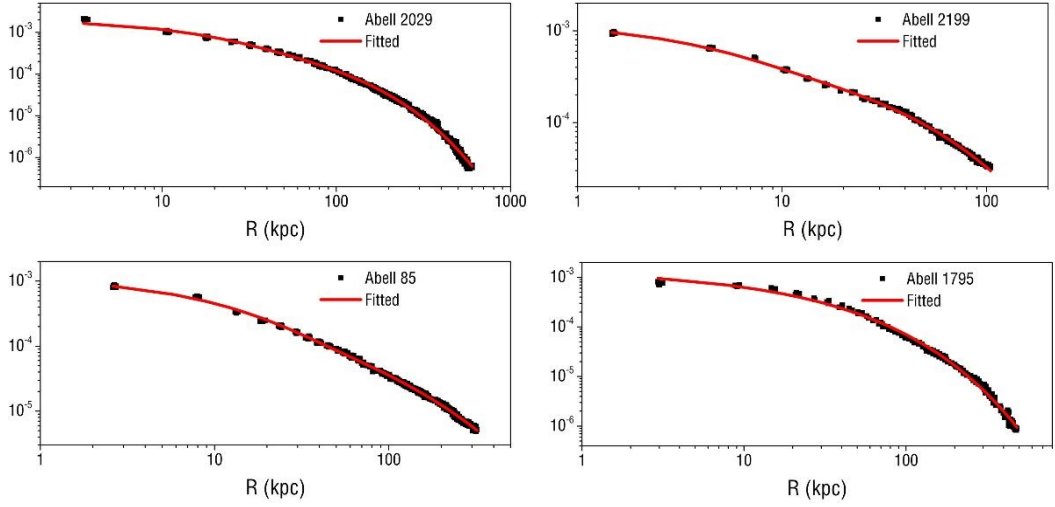

Fig. 2 Experimental surface brightness data from actual observations and the fitted curves. We analyzed the surface brightness profiles of four galaxies, Abell 2029, Abell 2199, Abell 1413, and Abell 1795, respectively. The unit of the y-axis is cls s$^{-1}$ arc sec$^{-2}$. To improve the overall fit, two stretched exponential functions were used in the fitting to capture both the short-distance decay and the intermediate-to-long distance. The fit parameters are listed in Table 1.

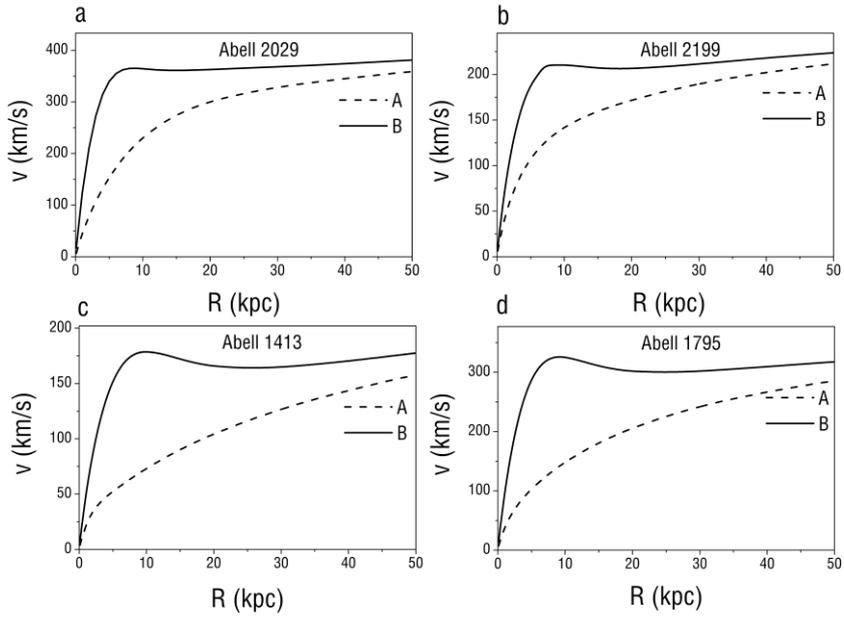

Fig. 3 The calculated rotation curves for four different galaxies. Curves A in dash line, were calculated using Eqs. (7) and (8), solely based on Newtonian dynamics and the fitted parameters in Table 1 for the experimental surface brightness profile. All dash-line curves exhibit an initial rise in the velocity with an asymptotic flat top. In contrast, curves B in solid line are calculated by including an extra contribution from the galactic bulge with a density given by Eq. (9). All red color curves of the B-type show a steeper rise to a flat top, and they more closely resemble the actual rotation curves.

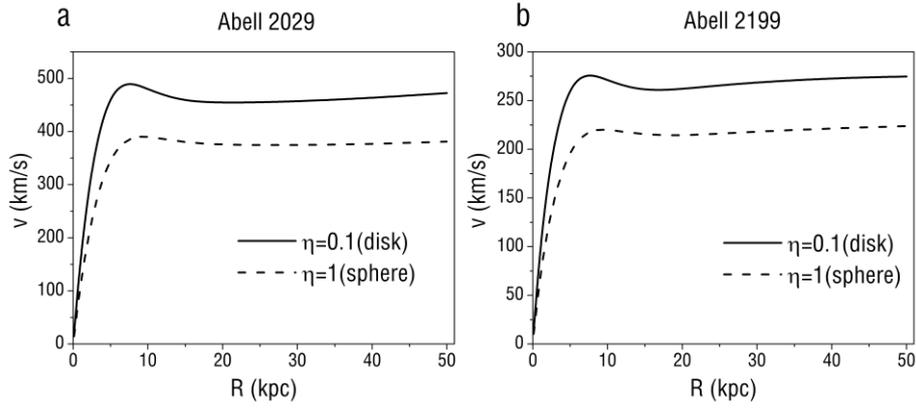

Fig. 4 The calculated rotation curves based on Eq. (14) for Abell 2029 and Abell 2199, assuming an ellipsoid-shaped density profile.   The flatness parameter $\eta$ of the ellipsoid can influence the shape and the value of the rotation curve.   In comparison, the rotation velocity increases as $\eta$ decreases.   The curve for a spherical density profile with $\eta = 1$, also shown previously in Fig. 3, has a lower velocity than that of a disk with $\eta = 0.1$, a typical value for spiral galaxies.